# Innovative Software Development and Project Management Framework for Technology Startups

**SONG-KYOO KIM**

## ABSTRACT

This paper introduces a new process that integrates inventive problem-solving methods into modern software development. The central research question addresses how tech startups can enhance their software development processes with minimal project management expertise. The Systematic Innovation Mounted Software Development Process, which blends Agile and Systematic Innovation, offers an alternative framework to facilitate idea generation in software products. This intuitive project management framework empowers technology-driven companies to manage their projects more effectively. The study aims to provide guidelines for entrepreneurs to manage projects successfully. By collaborating with the existing Agile model, the Systematic Innovation model fosters creativity and innovative problem-solving. Ultimately, this new software development process and its techniques have the potential to significantly impact the software industry, particularly for startups, as they alleviate managerial burdens and allow companies to focus on their core technologies.

## 1. INTRODUCTION

In the modern era, a software is complicated and required for the innovation. The development process has evolved to cover the wide range of software applications. The paper provides the enhanced software development process for the recent software industry which is one of the most rapidly growing industry sectors. In addition, software development startup companies in this industry are the main economic foundations of most countries in the world (Wang and King, 2000). Because of the above reasons, improvement of the software development process within a short product lifecycle under low costs has been targeted to maintain minimum quality. A software process framework is required to build high quality software (Pressman, 2001) even though it is also well known that there is no silver bullet software development method that fits for all (Sommerville, 2011). Software development companies have their own cultures, characteristics and target market for their products. Therefore, one software development process cannot be utilized for all predefined current development processes. Companies must adapt development processes according to their own protocol based on needs and contexts. In addition, there are natural conflicts between delivery time, the cost and quality that impact the software development process (Pressman, 2001). The modern software process still strives to maintain a certain level of quality over efficiency. Some research indicates that repairing after software product launch could cost a hundred times more than removing the problem during the initial phase (Pressman, 2001). Therefore, software development processes need to be continuously refined and improved, and Software Process Improvement (SPI) could be one of practical solutions to achieve the software enhancement. SPI is a systematic procedure for enhancing the performance of the development process by changing the current process (Sommerville, 2011). It also includes driving the implementation of changes to that process to achieve specific goals such as increasing development speed, achieving higher product quality or reducing costs. Someone who leads SPI must understand the methodologies and the tools to be adapted within the current circumstance and understanding the state of practice and process improvement initiatives accordingly (Adolph, 2012). SPI has been applied to various areas of software development sectors. Electronic commerce software development has different requirements than a conventional system of software development (Gruhn and Schope, 2002). SPI can be applied to a highly integrative electronic commerce system development project, to avoid the danger of failing the entire project. SPI could include new activities in the development process, and removing some of them as well (Gruhn and Schope, 2002). Some researchers provide guidelines to engage software process improvement for small and medium companies by analyzing critical SPI requirements (Sulayman and et. al., 2014). Even though there is some research for SPI models targeted for small and medium companies, there are no dedicated SPI models for technology startup companies which have smaller numbers of employees who utilize a rapid software development cycle. In addition, the current SPI models are mostly targeted for large companies and the models are not suitable for small sized companies (less than ten persons) because these models are expensive and complicated (Kim, 2010). Recently, some researchers have contributed modified SPI models that apply to both small and large sized companies by a competitive advancement strategy (Cater-Steel, 2004). Lean has also been adopted into the software development process by using some main principles from the original Lean manufacturing process (Poppendieck, 2003) and it is suitable for the technology startup companies that developing software products. As literature studies show, SPI implementation is an effective approach for software development process enhancement. In this recent environment of rapid change and development, the software industry generally requests that the development process should be more flexible to reflect changes during developments. Even though software development favors the Lean process, recent trends in the software sector indicate that products should be more innovative and appealing for consumers. Uniqueness of the software sector keeps changing due to competitive business environments, the organizations and cultures. Basically, it evolves within different development environments and the development process should evolve accordingly, which translates to more flexibility and innovation. This paper proposes newly integrated frameworks by using systematic innovation for software development and its project management, which has never been introduced before. The lean and compact SIM-Process to be integrated on the existing agile development process has no side effects or overload. Consequently, the Systematic Innovation process is suitable for software startup companies because the companies are small and lean.

In this research, it is tried to presents a combination of the systematic innovation with software development process that is suitable for dynamic development environments. The proposed development process (Systematic Innovation Mounted Process) is conceptually similar with the Agile and the Lean process but it is designed for the innovative activities including the story producing and the user experience (UX) design. After explaining the Systematic Innovation Mounted Process, two study cases (Com2Us mobile game, web service developments) for software developments are introduced for showing how this new process and the project management framework could be applied into technology based startup companies. The main research method is a qualitative analysis based on the related documents and the interviews with professionals in this area.

## 2. THEORETICAL FRAMEWORK AND RELATED LITERATURES

The software development process is often considered as a subset of a systems development lifecycle (SDLC) for developing software products. There are several models for processes that describe approaches to a variety of tasks and activities that take place during the procedure.

Generally, the system development lifecycle is a broader term and the software development process is a more specific term. The software lifecycle typically includes the following steps: requirements, analysis, design, construction (or coding), testing (validation), installation, operation, and maintenance (Cohen and et. al, 2010). The international standard for software lifecycle (ISO, 2017) has mentioned that many software development processes fit the spiral lifecycle model from the system development lifecycle model. Agile software development is also adapting the spiral cycles (recursive, iterative) for enhancing the development process. Agile software development, based on iterative and incremental development, is practical for startup companies. Agile development processes were introduced in the 1990s, to minimize a process bureaucracy by removing unnecessary milestones because of the administrative workloads (Conboy and Morgan, 2011). Agile software development is targeted to deliver a software product quickly to consumers, who could also propose new business requirements into products.

This philosophy behind agile methods is reflected in the Agile manifesto which values individuals and interactions, working software, customer collaboration and responding to change (Conboy and Morgan, 2011). The philosophy of Agile Software Development is core in the reality of current markets. The emergence of agile software processes attempts to deal with issues introduced by rapidly changing and unpredictable markets (Highsmith and et. al, 2001). Exploring each value helps us to understand the philosophy of the agile process and activities to apply the philosophy, to enhance software development--aligning with the latest volatile markets. Feature Driven Development (Stephen and et. al., 2002), Scrum (Cohn, 2009), Extreme Programming, Crystal, Dynamic System Development Method (Stapleton, 2003) and Adaptive Software Development (Highsmith, 2000) are common software development methodologies that are aligned with the values of Agile Software Development (AL-Taani and Razali, 2013). In the view of organizations, agile development activities are suitable for small, co-located, dedicated and highly collaborative teams (Boehm, 2002; Nerur, et. al., 2005).

In general, agile development is regarded as the extreme opposite of Waterfall development. In the Agile process, a series of these processes are repeated, known as reputation development (Furugaki, 2007). There are some variances of the Agile development processes, which usually starts with *Planning* phase and defines *Requirements, Design, Implementation*, *Testing (and Integration)* and *Evaluation* (see Figure 1) phases. It has several recursive cycles (iterations) between *Design* and *Testing and Integration* phases. Once all requirements are determined as actual implementation during the Evaluation phase, the project is completed (Product phase). The recursive cycle (*Design; Implementation; Testing and Integration*) is a minimum set of the development cycle, executed daily. A Daily-based Scum framework is commonly applied for this recursive cycle (Cohn, 2009). Scrum is a flexible and holistic development strategy that is an iterative and incremental agile software development framework where a development team works as one unit to reach a common goal (Cohn, 2009).

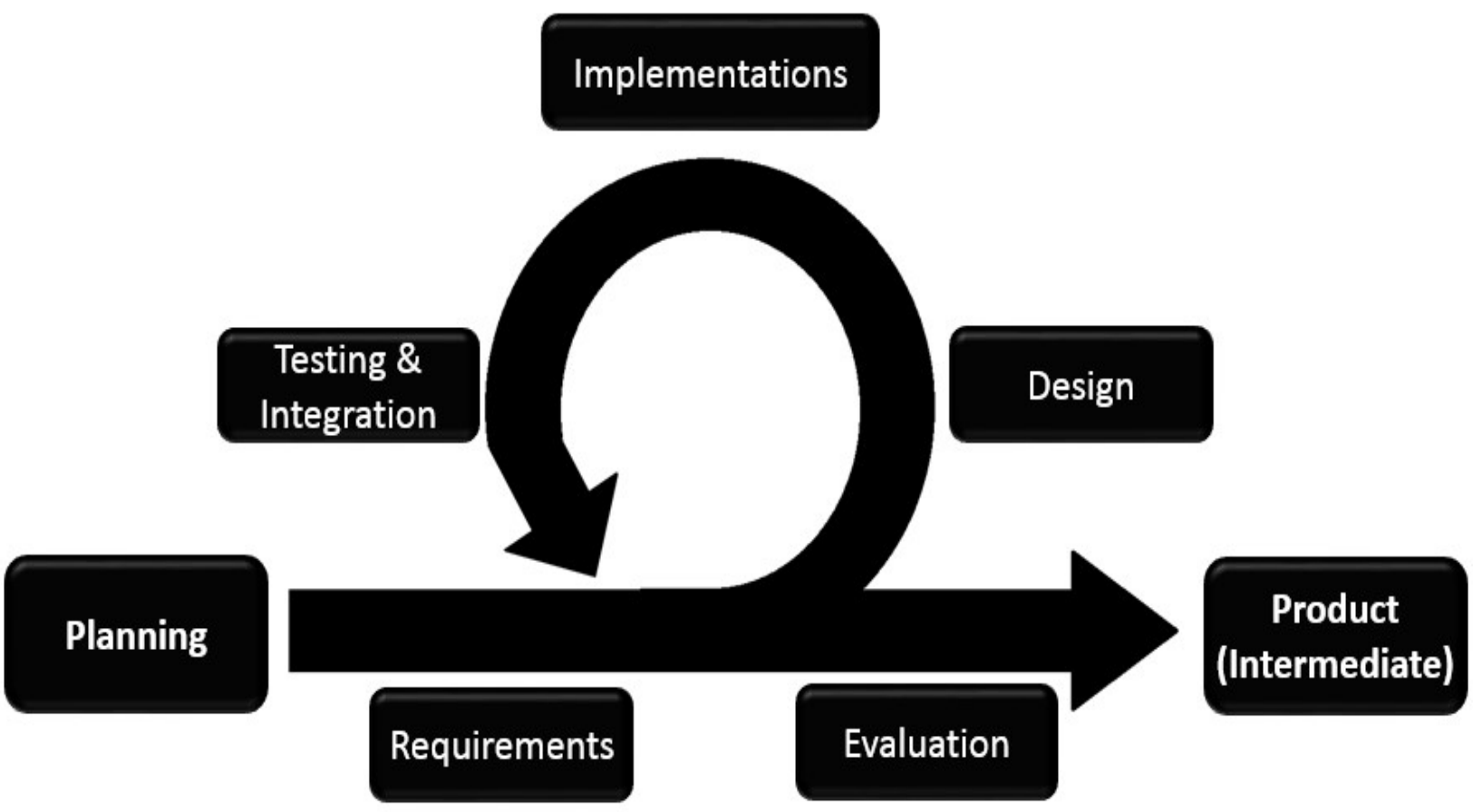

**Figure 1.** Agile software development process

Systematic Innovations (SI) is a structured process and the set of practical tools for new idea generations and application to technical problems, including software implementation issues (Terninko and et. al, 1998). The tools of Systematic Innovation have been widely used for technical breakthroughs and system improvements (Petkovic and et. al, 2013).

In general, problem solving and innovation processes which include 6-sigma, Lean Thinking, IDEO process, ARIZ (Algorithm of Inventive Problem Solving) and SI (Systematic Innovation) usually contain one or several task blocks (also called phases) to generate and implement new ideas and solutions. This phase approach of innovation method provides check points to use inventive problem solving tools more effectively. An 8-step phase approach is widely used in the systematic innovation process (C2C Solutions, 2016), however a 3-step phase process is used for mounting with the existing Agile processes in this research. This 3-step process for Systematic Innovation has recently been introduced for those who use complicated problem solving tools with ease. This 3-step process is simple and easy to use, even for beginners. A detailed description of each step for Systematic Innovation is as follows, adopted from Kim (2015):

***Step 1: Problem Identification:***

Throughout this step, users identify shortcomings in their idea generation and problem solving capabilities. This step identifies the What-I-Want (WIW) that is key for formulating the problem. This step is similar with the value identification in Lean Thinking (Womack and Jones, 1996). ENV (Element-Name-Value) model in OTSM (General Theory of Powerful Thinking)-TRIZ (Khomenko and Guio, 2015) and RCA (Root Cause Analysis) are typical inventive problem solving (TRIZ) tools used during this step. In some business development problems, a good definition may lead to immediate identification of possible solutions. This step acts as a preliminary process for making the problem simple and clear, through the use of several systematic innovation tools.

***Step 2: Problem Solving:***

The problem-solving step moves to generate the concept solution starts from the formulated problem during the problem identification step. Most of TRIZ tools, such as the 40 Inventive Principles, Substance-Field model with 76 Standards (Domb, 2003) and ARIZ (Altshuller, 1989), are applied in the Problems Solving step. While the tools are mostly adapted from the TRIZ method, a user could adapt tools from other methods such as Lean Thinking and Six-Sigma.

***Step 3: Evaluation and Prototyping (Concept Design and Evaluation)***

This step helps idea generators to choose the most suitable solutions for implementation from among numerous possible solutions generated. Selection of the candidate solutions and actual implementation are in this last step. Based on the concept solutions, users can develop prototype solution to be applied in the problem situations.

The Systematic Innovation method can be iterative as it is a set of continuous evolving tools that improve the ability to solve problems. TRIZ (TIPS; Theory of Inventive Problem Solving) is the most powerful tool set for Systematic Innovations (Altshuller, 1996; Domb, 1999; Grace and et. al., 2001). In addition, the SI methods can be easily collaborated with Lean Thinking and Six-Sigma activities. For instance, tools in the *Problem Identification* phase can replace tools in the Value Identification phase during Lean Thinking activities (Womack and Jones, 1996), which is similar to the first phase of the agile development process. Tools in the *Problem-Solving* phase can also replace the tools for the Design (Optimize) phase in Design for Six Sigma (DFSS) activities (Breyfogle, 1999) (see Figure 2). Systematic Innovation is originally targeted to solve engineering problems but the method has expanded to various areas included in new software development (Kim, 2010; 2011; 2012; 2016).

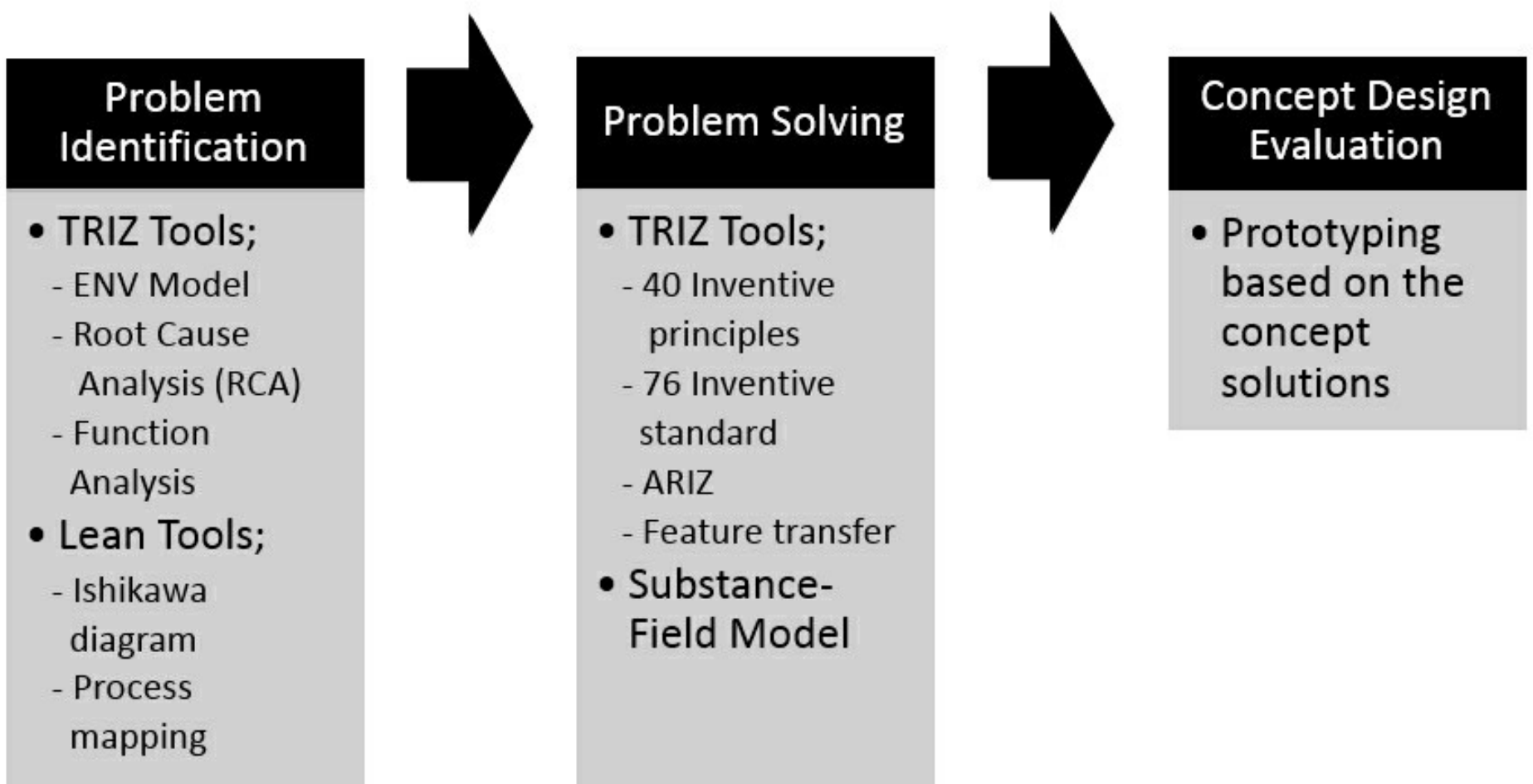

**Figure 2.** Systematic Innovation process

The Agile process introduces the idea of simplicity. The more effort needed to find required information, the more effort is needed to keep the information up to date (Nokes and et. al., 2007). Agile software development fits well in terms of flexibility to reflect the requirements from astute customers. The innovative properties of software products have become a mandatory factor for success in business under uncertain circumstances. Software companies need to think differently to generate new ideas that appeal to customers. Adapting distinct innovation processes that can be widely used by world-leading companies could be one way to move for delivering innovation into products. DeepDive (IDEO), Lead User Research (3M) and Design Thinking (Apple) are well known innovation processes that have been adapted and are being used by various companies. Integrating an innovation process into an existing software development process is not an easy task. The innovation process must be lean enough to avoid associated side effects, confusion, and overload during the integration. Consequently, the innovation processes mentioned above might be not applicable for software startup companies due to characteristics of the companies, which are small and lean.

The project management role is vital and project management for software development is also a critical factor for success. Project management is the process and activity of planning, organizing, motivating, and controlling resources, procedures, and protocols to achieve specific goals in scientific or daily problems (Meredith, 2011). A project is a temporary endeavor designed to produce a unique product, service or result with a defined beginning and end (usually time-constrained, and often constrained by funding or deliverables) (Nokes and et. al., 2007), undertaken to meet unique goals and objectives, typically to bring about beneficial change or added value (Dinsmore and et. al., 2005).

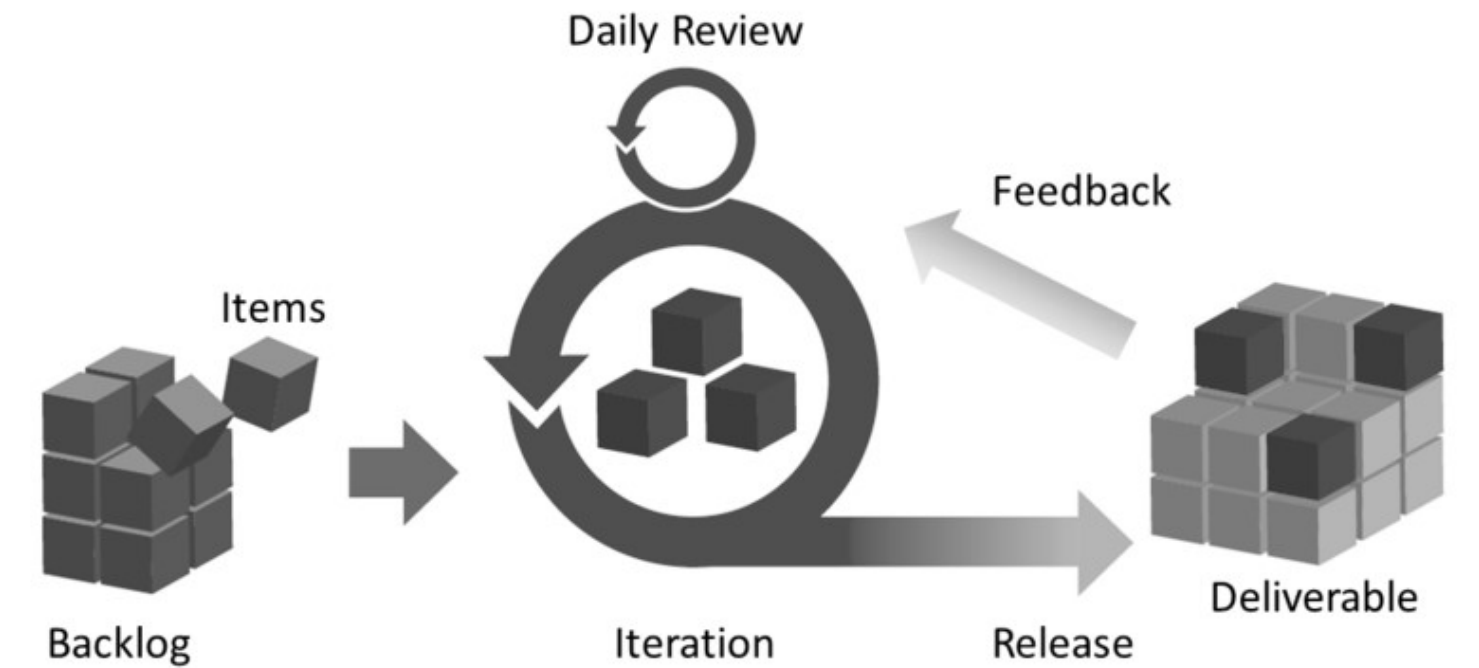

**Figure 4.** Agile project management diagram from Planbox (2012)

## 3. PROPOSED METHODS FOR SOFTWARE DEVELOPMENT

### 3.1 Systematic Innovation Mounted Software Development Process

Systematic Innovation can be adapted to the agile development process to implement the software innovatively. Systematic Innovation Mounted (software development) Process (SIM-Process) is a framework to help developers to generate more innovative ideas systematically. Some steps in the software development are required to create something new and it requires to use the innovative problem solving skills. The SIM-Process could be adopted for these type of steps including the (story) idea generation and the graphic design especially for mobile game developments.

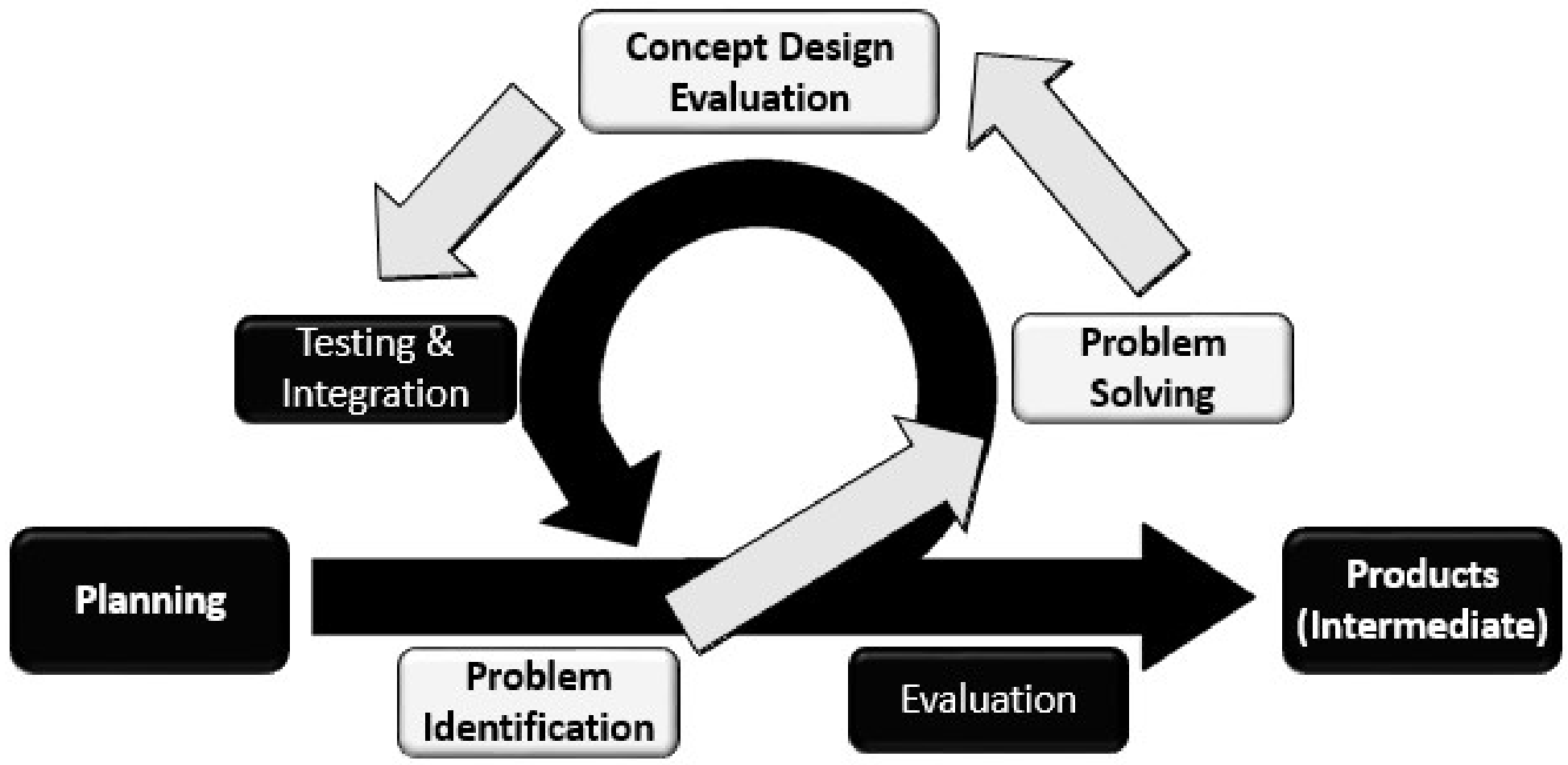

**Figure 3.** Systematic Innovation Mounted software development process

The general procedure of a SIM-Process is like an Agile software development processes except for some phases in the agile process. *Requirements, Design,* and *Implementation* phases in the agile process are replaced with *Problem Identification, Problem Solving,* and *Concept Design* phases in the SI method (see Figure 3). Sometimes, software development requires to be generate new idea innovatively. The SIM-Process is suitable for developing the new idea in the different perspective and it is really handy to be attached with atypical software development process like the Agile process. Beside of the Systematic Innovation, other components of the software development are the same as an existing software process. The SIM-Process could be adopted under the exact same setups including the setups of human resources, time managements for daily scrum, the release and sprint plans, and development roadmap. The SIM-based software development is newly introduced in this paper, but the Systematic Innovation process has been widely used for high level feature design and user experience design in the software development (Kim, 2010; 2011; 2012; 2016). It is very practical for high level conceptual programming that mostly requires innovative ideas but it is rarely applied in practice because there is no integrated framework to merge various concepts and tools at once.

### 3.2. Intuitive Project Management Framework

The new framework in this paper suggests modules for project management, and the whole project consists of building blocks as unit modules. Unlike existing project management, the resources of each task such as duration of the task, number of human resources and costs are defined into the unit module. So, the project manager can determine workloads by counting number of the unit modules in the project diagram. One unit module is either an Agile process which contains a whole cycle of the development process or a SIM-Process which contains a whole process for idea generations. The resource usage (duration, human resource, and cost) of one SIM-Process module is assumed to be the same as what an applied agile process module is (see Figure 4). For instance, the project manager determines one agile process module of five daily iterations as one unit, which means one unit module is completed within one week (assume one week as five working days; $\alpha$=5). Sometimes the project manager needs to split the project into several sub-projects (splitter) and vice versa (integrator). The project manager may need to reconsider the whole project even if in the middle of the development (checker) phase because of changing requirements. This set of new project management planning requirements is built, based on various circumstances of the software development process (see Table 1). Each module is one single block and a project manager can build up the software development process by adding these modules (see Table 1).

| Module | Symbol | Text Symbol | Variable | Descriptions |
|---|---|---|---|---|
| Agile Process | x y | # | X: one recursive (iteration) cycle<br>Y: process duration | Module of Agile software development process |
| SI Mounted Process | x y | * | X: one recursive (iteration) cycle<br>Y: process duration | Module of Systematic Innovation mounted software development process |
| Integrator | Z(X, n) | } | Z: integration duration depend on nodes | Module for Integration of the development modules |
| Splitter | x | { | X: split duration | |
| Checker | C x | C | X: check durations | Checksum and reconsidering the whole project (the project could be terminated). |
| Start/End | Start End | @ | N/A | Start and end of the project |

**Table 1.** Module set of software development project

## 4. COM2US: ACTION PUZZLE FAMILY

Com2uS has been a leader of the mobile game industry since its inception in 1998 and the company built its reputation as the number one mobile games provider in South Korea (Kim, 2014). Com2uS was a successful developer of many premium titles. The name of Com2uS remains as the leader in the mobile game industry even though the company merged with Gamevil in 2013. Action Puzzle Family (APF) is in the form of delightful classical easy puzzle games. This game tittle has been known as one of the popular Freemium casual games, which has more than six million users of eight puzzle games [26]. The goal of the game is collecting all puzzle pieces to move the family into a house with wacky features and each one of the eight eccentric puzzle games has a different theme. The APF game project started in June, 2006 with 10 members, including a producer and software developers that lasted until the end of the project, August 2007. The APF was developed by a small group (less than ten people in total) and the process was flexible than the general Com2uS development process. According to Com2uS case research [26], idea generation and graphic design phases are required for generating the new and fresh ideas; the SIM-Process module might be suitable for this process. In the general Com2uS development process, bug-fix tasks before starting alpha and beta testing phases are mandatory. The APF project could be described by using the intuitive project management framework more effectively and two steps are required for creative problem-solving skills. So, the SIM-Process is applied into these two steps which are the Idea Generation and the Graphic Design (see Fig. 4).

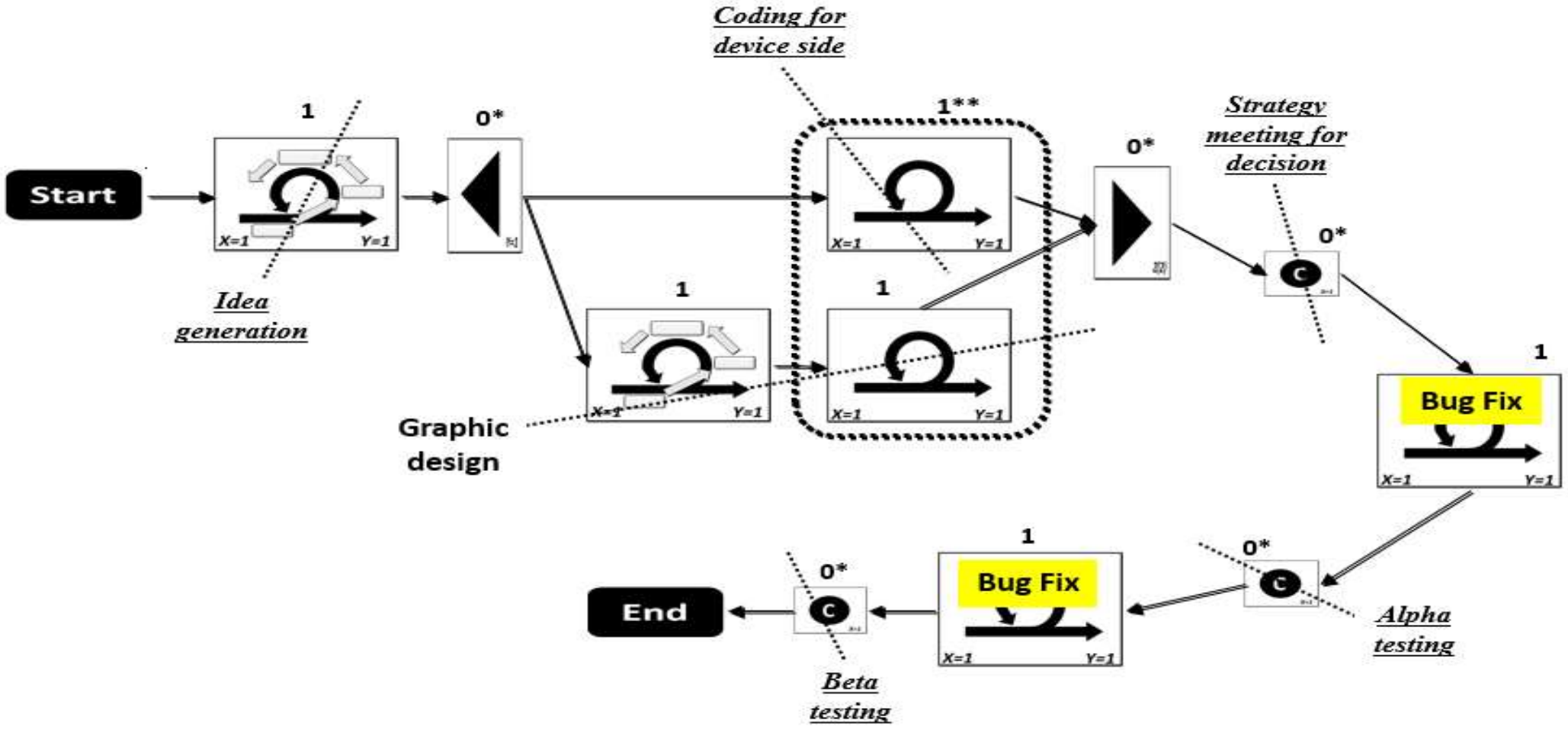

*) Based on the pre-defined functions (Table 2),
**) Assume the human resources are not limited

**Figure 4.** Institutive AFP project planning diagram

The APF producer could determine the unit of modules and four parameters (X, Y, Z and α) during the project initiation. The values of the parameters (X, Y, Z and α) are as follows:

- X = 1 [week]
- Y = 1 [month]
- α = 4 [weeks/month].

The duration of the *integrator* before the *strategic decision meeting* is calculated as follows:

$$Z(X,\alpha,n) = Z(1,4,2) = Round\left(1 \cdot \frac{2 \cdot (2-1)}{2 \cdot 4}\right) = 0$$

and duration of the integration completion is less than one month (0*; 1 week). Three *checkers* of the project include, the moment to check if the APF project is worthy enough to complete based on the current plan (the project could be dropped at this moment). The APF project planning was also assumed that each member in the project was a highly-skilled programmer or engineer and each task may have different members. The APF project diagram (Figure 8) instantly shows that the project would takes 6 months and 1 week with 14 developers (2 members 7 exclusive tasks) by simply counting the number of 1s and 0*s (5 months and 5 weeks = 1+0*+1+1+0*+0*+1+0*+1+0*). The duration of all modules is calculated based on the predefined functions in Table 2. The APF project case has been used to demonstrate how the SIM-Process could be applied into the mobile game software. It is noted that the SIM-Process could be more practical when it is applied into simpler software development. The next case is a SIM software development application for another software development process which is simpler than the APF project.

| Variable | Meaning | Unit | Default Function |
|---|---|---|---|
| $X$ | Duration of recursive period (Unit period) | [day], [week] | N/A, constant |
| $Y$ | Duration for completion of one module process | [week], [month], [year] | $Y(X,\alpha) = Round\left(\frac{X}{\alpha}\right)$ |
| $Z$ | Duration of integrating the module | [week], [month], [year] | $Z(X,\alpha,n) = Round\left(X \cdot \frac{n(n-1)}{2 \cdot \alpha}\right)$ |
| $n$ | Number of nodes for integration | N/A | N/A |
| $\alpha$ | Duration unit between the cycle and module (i.e., number of cycles for one module process). | N/A | $\alpha = \frac{5[day]}{1[week]} = 5$ or $\alpha = \frac{4[week]}{1[month]} = 4$ |

**Table 2.** Parameters and pre-defined functions for the module set

Even though technology driven startup companies may begin as a small group which has less than ten persons, project management is still one of the mandatory skills for business success. In a typical startup company, one person might have multiple roles as a developer, project manager and marketer at the same time. These companies usually do not have enough human resources for assigning project management roles independently, however effective management of projects is as critical as product development. This framework provides the building blocks for intuitive project planning, especially for software development projects. It provides the flexibilities of innovative software development by using the SIM-Process.

## 5. FINDING AND RESULTS

The software development process is becoming more complicated because customers want more innovative and attractive products. The paper has provided a new process which integrates an inventive problem solving method into one modern software development program, making it part of the software development process. The implication of this research provides the guideline to help the entrepreneurs for managing their project properly because the Systematic Innovation model helps to generate new ideas and innovative ways to solve problems. The SIM-Process provides alternative approaches to create new ideas for adapting to innovative software products. In addition, the new project management framework helps technology driven companies to manage their projects intuitively. Unfortunately, to apply this new framework, each individual member requires having enough knowledge of either one of processes (Agile or Systematic Innovation) and project leaders might be advised to understand the minimum skill sets for project management. Otherwise, each module in the project may not be completed on time, which could lead to project delays.

## 6. DISCUSSION AND CONCLUSION

The practical cases, which could adopt the SIM-Process and the intuitive project management framework, demonstrate how these innovative methods can be applied into real project management scenarios. Actual adaptation of the SIM-Process with managing a whole project by using the institutive framework could be future research implication for the practices. The practical techniques in this paper could impact the current software development industry significantly, especially for software startup companies, because these powerful weapons could effectively lean down the managerial workloads of companies and make them stay focused on their core assets.